\pdfoutput=1
\documentclass[prl,twocolumn,aps]{revtex4}
\usepackage{graphicx,graphics,color,epsfig}% Include figure files
\usepackage{bm}
\usepackage{amsmath}
\usepackage{amssymb}
\usepackage{epstopdf}

\makeatletter

\newcommand{\Rmnum}[1]{\expandafter\@slowromancap\romannumeral #1@}
\newcommand*{\rom}[1]{\expandafter\@slowromancap\romannumeral #1@}
\makeatother

\usepackage{color}

\begin{document}

\title{Possible pairing symmetry in the FeSe-based superconductors determined by quasiparticle interference}

\author{Yi Gao,$^{1,2}$ Yuting Wang,$^{1}$ Tao Zhou,$^{3}$ Huaixiang Huang,$^{4}$ and Qiang-Hua Wang$^{5,6}$}
\affiliation{$^{1}$Center for Quantum Transport and Thermal Energy Science, School of Physics and Technology, Nanjing Normal University, Nanjing 210023, China\\
$^{2}$Jiangsu Key Lab on Opto-Electronic Technology, School of Physics and Technology, Nanjing Normal University, Nanjing 210023, China\\
$^{3}$Guangdong Provincial Key Laboratory of Quantum Engineering and Quantum Materials,
and School of Physics and Telecommunication Engineering,
South China Normal University, Guangzhou 510006, China\\
$^{4}$Department of Physics, Shanghai University, Shanghai, 200444, China\\
$^{5}$National Laboratory of Solid State Microstructures $\&$ School of Physics, Nanjing
University, Nanjing, 210093, China\\
$^{6}$Collaborative Innovation Center of Advanced Microstructures, Nanjing 210093, China}

\begin{abstract}
We study the momentum-integrated quasiparticle interference (QPI) in the FeSe-based superconductors. This method was recently proposed theoretically and has been applied to determine the pairing symmetry in these materials experimentally. Our findings suggest that, if the incipient bands and the superconducting (SC) pairing on them are taken into consideration, then the experimentally measured bound states and momentum-integrated QPI can be well fitted, even if the SC order parameter does not change sign on the Fermi surfaces. Therefore, we offer an alternative explanation to the experimental data, calling for more careful identification of the pairing symmetry that is important for the pairing mechanism.
\end{abstract}

\maketitle

%\section{introduction}

The superconducting (SC) mechanism and pairing symmetry in the FeSe-based superconductors, \emph{e.g.} A$_x$Fe$_{2-y}$Se$_2$ (A=Rb, Cs, K) \cite{chenxl,chenxh2,conder}, Li$_{1-x}$Fe$_x$OHFe$_{1-y}$Se \cite{chenxh0,chenxh,Johrendt,zhaozx,clarke0}, Li$_x$(NH$_2$)$_y$(NH$_3$)$_{1-y}$Fe$_2$Se$_2$ \cite{clarke}, as well as monolayer FeSe grown on SrTiO$_{3}$ \cite{xueqk1}, remain hotly debated ever since their discovery. The hole bands sink below the Fermi level and become incipient in these materials while there are only electron-like Fermi surfaces, contrary to the electron- and hole-like ones in the usual iron pnictides \cite{zhouxj1,zhouxj2,fengdl1,fengdl2,shenzx1,zhouxj3,fengdl4,ding,zhouxj4,fengdl5,fengdl6,shenzx2}. However the transition temperature in these materials is the highest among all the iron pnictides, the reason of which is still unclear.

To resolve the SC mechanism, various pairing symmetries have been proposed, including the nodeless $d$-wave \cite{aoki,scalapino,leedh,balatsky,kontani}, sign-preserving $s$-wave \cite{zhou,kontani,Fernandes}, hidden $s_\pm$-wave \cite{hu,wang,kontani,linscheid,mishra} and in$\&$out $s_\pm$-wave \cite{mazin2,chubukov}. Among them, the nodeless $d$- and in$\&$out $s_\pm$-wave symmetries show apparent sign reversal of the SC order parameter ($\Delta_\mathbf{k}$) on the Fermi surfaces, while the sign-preserving $s$- and hidden $s_\pm$-wave symmetries exhibit no such sign reversal. However for the hidden $s_\pm$-wave symmetry, there is a hidden sign change of $\Delta_\mathbf{k}$ between the incipient bands and the electron bands which cross the Fermi level.

Numerous experiments have been performed to distinguish the pairing symmetries. The SC gap magnitude measured by angle-resolved photoemission spectroscopy (ARPES) \cite{zhouxj1,shenzx1,zhouxj3,fengdl4,zhouxj4,fengdl5,fengdl6,shenzx2}, the density of states (DOS) measured by scanning tunneling microscopy (STM) \cite{xueqk1,xueqk4,fengdl3,xueqk5,fengdl7,wen}, as well as the temperature dependence of the London penetration depth \cite{yuan}, all suggest a nodeless SC gap, thus the nodeless $d$-wave symmetry seems to be ruled out since it would be nodal in the realistic Brillouin zone (BZ) where the Fermi surface warps along $z$ \cite{mazin2}. Inelastic neutron scattering (INS) has observed a spin resonance, which is interpreted as a sign-reversing $\Delta_\mathbf{k}$ on the Fermi surfaces \cite{Boothroyd,zhaoj1,Boothroyd1,Inosov1,Inosov2,Boothroyd2,Inosov3,zhaoj}. The in-gap bound states induced by nonmagnetic impurities, which are usually believed to indicate a sign-changing $\Delta_\mathbf{k}$ on the Fermi surfaces, have been observed in Ref. \cite{wen}, but not in Refs. \cite{fengdl3} and \cite{fengdl7}, therefore the former claimed that $\Delta_\mathbf{k}$ must change sign on the Fermi surfaces while the latter reached the opposite conclusion.

Furthermore, by measuring the quasiparticle interference (QPI) in the presence of magnetic vortices, Refs. \cite{fengdl3} and \cite{fengdl7} claimed a sign-preserving $s$-wave symmetry. However recently, Refs. \cite{hirschfeld1} and \cite{hirschfeld2} pointed out that the above conclusion may be model dependent and unreliable. Instead Hirschfeld, Altenfeld, Eremin, and Mazin proposed a so called HAEM method to process the QPI data and this method has been applied to bulk FeSe \cite{davis} and Li$_{1-x}$Fe$_x$OHFe$_{1-y}$Se \cite{wen}. Based on this method, Ref. \cite{wen} implied a sign-reversing $\Delta_\mathbf{k}$ on the Fermi surfaces.

In this work, we show that, when the incipient bands are present, nonmagnetic impurity-induced in-gap bound states can appear even if $\Delta_\mathbf{k}$ does not change sign on the Fermi surfaces. In addition, the quantity based on the HAEM method shows similar behavior between the hidden $s_\pm$- and in$\&$out $s_\pm$-wave symmetries. Therefore, we offered an alternative explanation to the pairing symmetry drawn from the QPI measurement in Ref. \cite{wen}.

\begin{figure}
\includegraphics[width=1\linewidth]{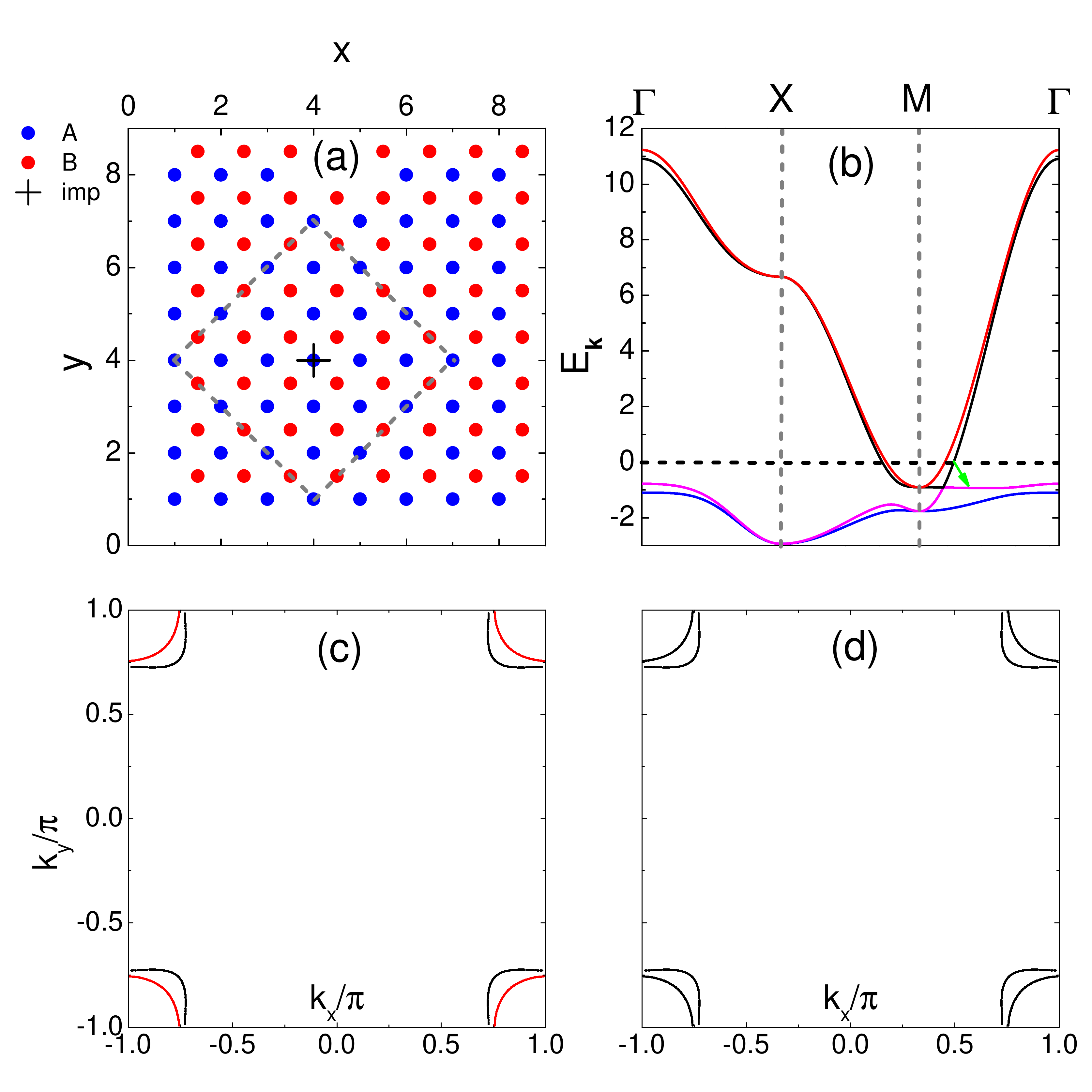}
 \caption{\label{position} (color online) (a) The iron lattice. The blue and red dots represent the $A$ and $B$ sublattices, respectively. The black cross denotes the position of the impurity, while the gray dashed square indicates the area we use to perform the Fourier transformation. (b) Calculated band structure along the high-symmetry directions in the 2Fe/cell BZ. The energy is defined with respect to the Fermi level (the black dashed line). The green arrow denotes schematically an off-shell scattering process that contributes to $\delta\rho^-(\omega)$ in the hidden $s_{\pm}$ pairing state. (c) The normal-state Fermi surfaces and the sign of the SC order parameter on them, for the in$\&$out $s_{\pm}$ pairing. Here the black and red color indicates that the order parameter is positive and negative, respectively. (d) is similar to (c), but is for the hidden $s_{\pm}$ pairing.}
\end{figure}

%\section{method}
We adopt a two-dimensional tight-binding model of the iron lattice, where each unit cell accommodates two inequivalent sublattices $A$ and $B$ [see Fig. \ref{position}(a)]. The coordinate of the sublattice $A$ in the unit cell $(i,j)$ is $\mathbf{r}_{ij}=(i,j)$ while that for the sublattice $B$ is $\mathbf{r}_{ij}+\mathbf{d}$, with $\mathbf{d}$ being $(0.5,0.5)$. Here we have taken $\sqrt{2}a$ as the length unit, where $a$ is the distance between the nearest-neighbor iron atoms. The Hamiltonian can be written as $H=\sum_{\mathbf{k}}\psi_{\mathbf{k}}^{\dag}A_{\mathbf{k}}\psi_{\mathbf{k}}$, where $\psi_{\mathbf{k}}^{\dag}=(c_{\mathbf{k}A1\uparrow}^{\dag},c_{\mathbf{k}B1\uparrow}^{\dag},c_{\mathbf{k}A2\uparrow}^{\dag},c_{\mathbf{k}B2\uparrow}^{\dag},c_{-\mathbf{k}A1\downarrow},c_{-\mathbf{k}B1\downarrow},c_{-\mathbf{k}A2\downarrow},c_{-\mathbf{k}B2\downarrow})$ and
\begin{eqnarray}
\label{h}
%\psi_{\mathbf{k}}^{\dag}&=&(c_{\mathbf{k}A1\uparrow}^{\dag},c_{\mathbf{k}B1\uparrow}^{\dag},c_{\mathbf{k}A2\uparrow}^{\dag},c_{\mathbf{k}B2\uparrow}^{\dag},\nonumber\\
%&&c_{-\mathbf{k}A1\downarrow},c_{-\mathbf{k}B1\downarrow},c_{-\mathbf{k}A2\downarrow},c_{-\mathbf{k}B2\downarrow}),\nonumber\\
A_{\mathbf{k}}&=&\begin{pmatrix}
M_{\mathbf{k}}&D_{\mathbf{k}}\\D_{\mathbf{k}}^{\dag}&-M_{-\mathbf{k}}^{T}
\end{pmatrix},\nonumber\\
M_{\mathbf{k}}&=&\begin{pmatrix}
\epsilon_{A,\mathbf{k}}&\epsilon_{T1,\mathbf{k}}&\epsilon_{xy,\mathbf{k}}&0\\
\epsilon_{T1,\mathbf{k}}^{*}&\epsilon_{B,\mathbf{k}}&0&\epsilon_{xy,\mathbf{k}}\\
\epsilon_{xy,\mathbf{k}}&0&\epsilon_{A,\mathbf{k}}&\epsilon_{T2,\mathbf{k}}\\
0&\epsilon_{xy,\mathbf{k}}&\epsilon_{T2,\mathbf{k}}^{*}&\epsilon_{B,\mathbf{k}}
\end{pmatrix}.
\end{eqnarray}
Here $c_{\mathbf{k}A1\uparrow}^{\dag}/c_{\mathbf{k}A2\uparrow}^{\dag}$ creates a spin up electron with momentum $\mathbf{k}$ on the $d_{xz}/d_{yz}$ orbital of the sublattice $A$. $\epsilon_{A,\mathbf{k}}=-2(t_{3}\cos k_{x}+t_{4}\cos k_{y})-\mu$, $\epsilon_{B,\mathbf{k}}=-2(t_{3}\cos k_{y}+t_{4}\cos k_{x})-\mu$, $\epsilon_{xy,\mathbf{k}}=-2t_{5}(\cos k_{x}+\cos k_{y})$, $\epsilon_{T1,\mathbf{k}}=-t_{1}[1+e^{-i(k_{x}+k_{y})}]-t_{2}(e^{-ik_{x}}+e^{-ik_{y}})$ and $\epsilon_{T2,\mathbf{k}}=-t_{2}[1+e^{-i(k_{x}+k_{y})}]-t_{1}(e^{-ik_{x}}+e^{-ik_{y}})$. Throughout this work, the momentum $\mathbf{k}$ is defined in the 2Fe/cell BZ and the energies are in units of 0.1 eV. In the following we set $t_{1-5}=1.6,1.4,0.4,-2,0.04$ and $\mu=-1.8673$ to fit the band structure measured by ARPES. Under this set of parameters, the average electron number is $n\approx2.12$ (the system is about $12\%$ electron doped). The band structure and Fermi surfaces in the normal state are plotted in Figs. \ref{position}(b) and \ref{position}(c). The top of the incipient bands at $\Gamma$ and the bottom of the electron bands at $M$ are both located at about 80 meV below the Fermi level, while the Fermi momentum is $k_F/\pi\approx0.25$, agreeing qualitatively with the ARPES measurements \cite{zhouxj1,zhouxj3}. The band structure and the pairing function in the band basis can be obtained through a unitary transformation $Q_{\mathbf{k}}$ as
\begin{eqnarray}
\label{unitary}
Q_{\mathbf{k}}^{\dag}M_{\mathbf{k}}Q_{\mathbf{k}}&=&\begin{pmatrix}
E_{1\mathbf{k}}&0&0&0\\
0&E_{2\mathbf{k}}&0&0\\
0&0&E_{3\mathbf{k}}&0\\
0&0&0&E_{4\mathbf{k}}
\end{pmatrix},
%\Delta_{\mathbf{k}}&=&Q_{\mathbf{k}}^{\dag}D_{\mathbf{k}}Q_{-\mathbf{k}}^{*}=Q_{\mathbf{k}}^{\dag}D_{\mathbf{k}}Q_{\mathbf{k}}.
\end{eqnarray}
and $\Delta_{\mathbf{k}}=Q_{\mathbf{k}}^{\dag}D_{\mathbf{k}}Q_{-\mathbf{k}}^{*}=Q_{\mathbf{k}}^{\dag}D_{\mathbf{k}}Q_{\mathbf{k}}$. Here $E_{1\mathbf{k}},E_{2\mathbf{k}}$ are the energies of the two incipient bands and $E_{3\mathbf{k}},E_{4\mathbf{k}}$ are those of the two electron bands ($E_{1\mathbf{k}}\leq E_{2\mathbf{k}}\leq E_{3\mathbf{k}}\leq E_{4\mathbf{k}}$). The diagonal components in $\Delta_{\mathbf{k}}$ represent the pairing function on each band while the off-diagonal ones signify the inter-band pairing, which we ignore for simplicity.

\begin{figure}
\includegraphics[width=1\linewidth]{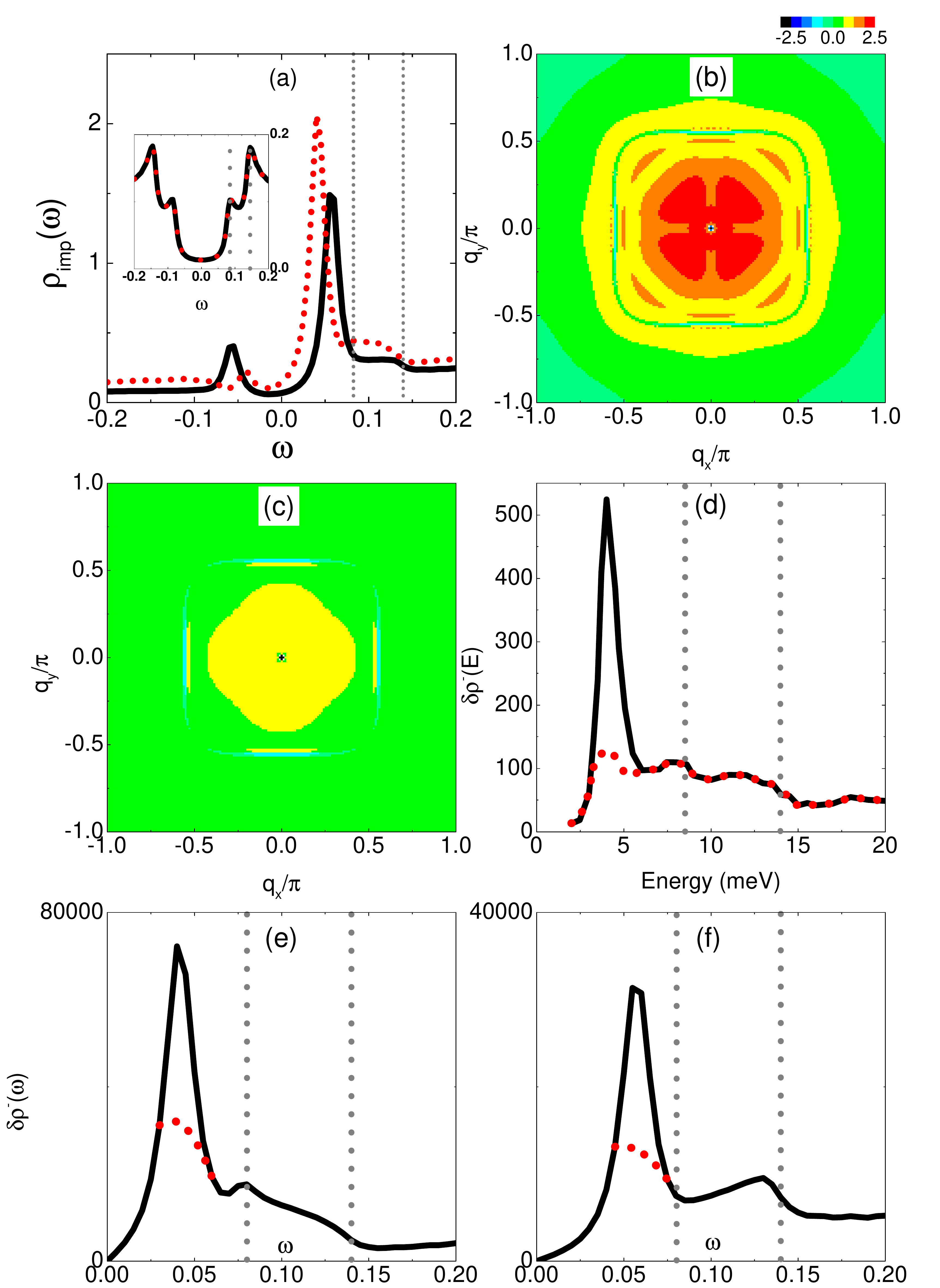}
 \caption{\label{dos} (color online) (a) The LDOS at the impurity site in the SC state. The red dotted curve is for the in$\&$out $s_{\pm}$ pairing ($V_1=6$, $V_2=-1$), while the black solid one is for the hidden $s_{\pm}$ pairing ($V_1=10$, $V_2=5$). The inset shows the DOS in the clean system. (b) The difference of the FT-QPI $\delta\rho^-(\mathbf{q},\omega=0.085)=\rm{Re}[\rho(\mathbf{q},\omega=0.085)-\rho(\mathbf{q},\omega=-0.085)]$, for the in$\&$out $s_{\pm}$ pairing. (c) is similar to (b), but is for the hidden $s_{\pm}$ pairing. (d) $\delta\rho^-(\omega)$ extracted from Figs. 3(b) and 3(c) of Ref. \cite{wen}. The black solid and red dotted curves are the original and filtered $\delta\rho^-(\omega)$, respectively. (e) and (f) are both similar to (d), but are our calculated results for the in$\&$out $s_{\pm}$ and the hidden $s_{\pm}$ pairings, respectively. The gray dotted lines in (a), (d), (e) and (f) indicate the position of the SC coherence peaks at $\omega=\Delta_1$ and $\Delta_2$.}
\end{figure}

\begin{figure}
\includegraphics[width=1\linewidth]{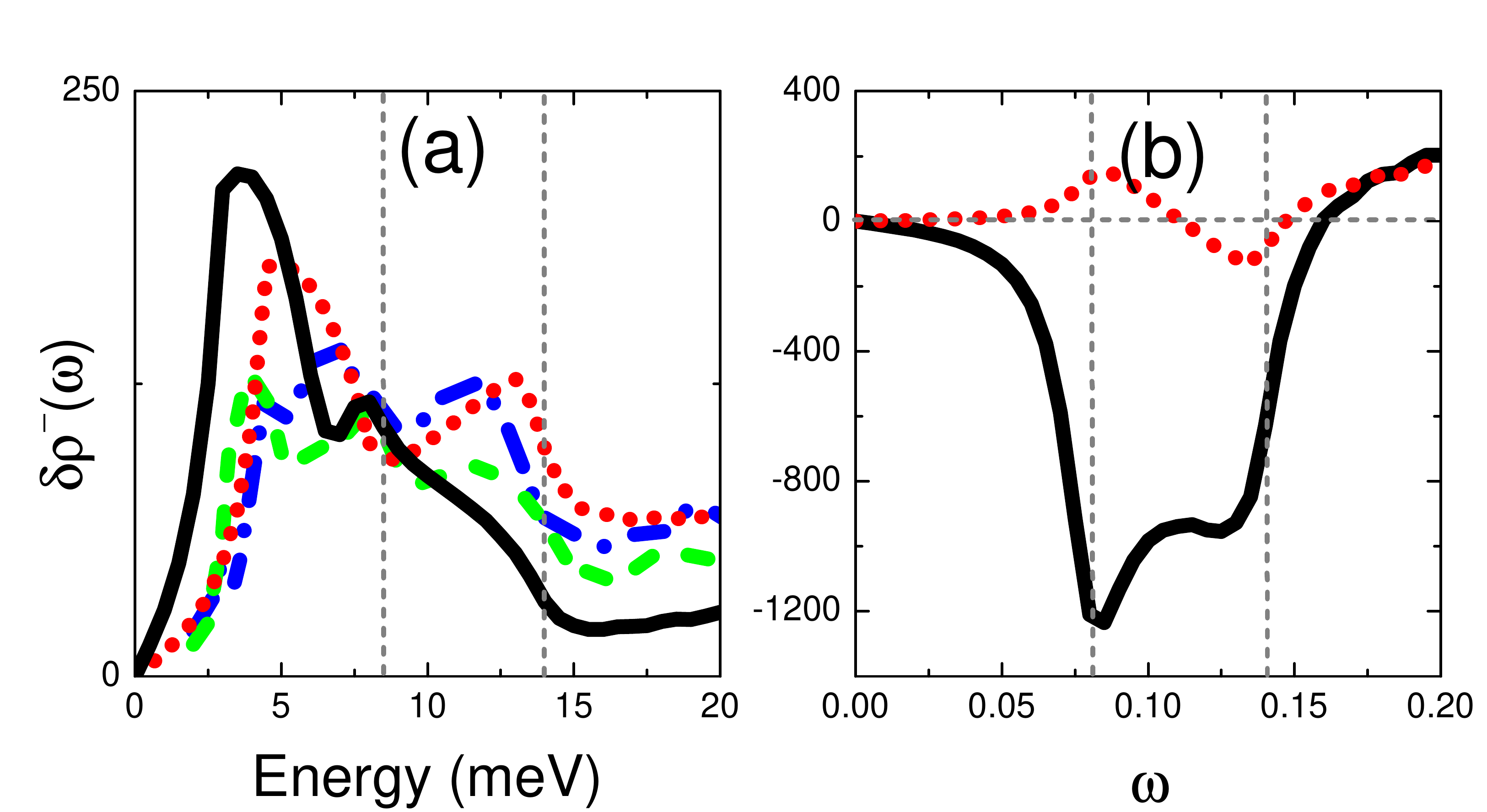}
 \caption{\label{comparison} (color online) (a) The filtered $\delta\rho^-(\omega)$. The black and red curves are our theoretical results for the in$\&$out $s_{\pm}$ and hidden $s_{\pm}$ pairings, respectively, while the green and blue ones are the experimental data extracted from Figs. 3(c) and S7(d) of Ref. \cite{wen}. (b) Theoretically calculated $\delta\rho^-(\omega)$ by ignoring the incipient bands. The black and red curves are for the in$\&$out $s_{\pm}$ and hidden $s_{\pm}$ pairings, respectively.}
\end{figure}

For the SC pairing, we consider two cases. The first one is the in$\&$out $s_{\pm}$ pairing, where we set
\begin{eqnarray}
\Delta_{\mathbf{k}}&=&\begin{pmatrix}
\Delta_2&0&0&0\\
0&\Delta_2&0&0\\
0&0&\Delta_2&0\\
0&0&0&-\Delta_1
\end{pmatrix},
\end{eqnarray}
with $\Delta_1=0.14$ and $\Delta_2=0.08$. It will lead to a sign-changing gap between the inner and outer electron pockets, as shown in Fig. \ref{position}(c). This pairing symmetry was suggested when the hybridization between the electron bands is strong enough \cite{mazin2,chubukov}. Another one is the hidden $s_{\pm}$ pairing, where we set
\begin{eqnarray}
\Delta_{\mathbf{k}}&=&\begin{pmatrix}
-\Delta_1&0&0&0\\
0&-\Delta_1&0&0\\
0&0&\Delta_2&0\\
0&0&0&\Delta_1
\end{pmatrix}.
\end{eqnarray}
Contrary to the in$\&$out $s_{\pm}$ pairing, the hidden $s_{\pm}$ pairing will not lead to any sign change of the gap along the Fermi surfaces, as shown in Fig. \ref{position}(d). However, the sign of the order parameter on the incipient bands is opposite to that on the electron bands. This pairing symmetry is predicted by the spin-fluctuation theory in the strong coupling limit \cite{linscheid,mishra}. In both cases, we have neglected the orbital selective renormalization effects \cite{davis,kreisel} by assuming a $\mathbf{k}$-independent $\Delta_{\mathbf{k}}$. A $\mathbf{k}$-dependent $\Delta_{\mathbf{k}}$ may affect the momentum dependence of the QPI signal, but since we are focusing on the momentum-integrated QPI signal in the following, we believe this assumption is reasonable and will not change the results qualitatively.
%Since the incipient bands are below the Fermi level and do not form Fermi surfaces, thus we denote this pairing symmetry as the hidden $s_{\pm}$ pairing.

For a single nonmagnetic impurity located at the $A$ sublattice of the unit cell $\mathbf{R}=(0,0)$, the impurity Hamiltonian can be expressed as
$H_{imp}=\sum_{\alpha,\beta=1}^{2}\sum_{\sigma=\uparrow,\downarrow}V_{\alpha\beta}c_{\mathbf{R}A\alpha\sigma}^{\dag}c_{\mathbf{R}A\beta\sigma}$. Since it is a multiorbital system, the scattering may consist of both the intraorbital ($V_{\alpha=\beta}=V_1$) and interorbital ($V_{\alpha\neq\beta}=V_2$) components. Following the standard $T$ matrix procedure \cite{zhu}, we can obtain $\rho_{A/B}(\mathbf{r}_{ij},\omega)$, which is the local density of states (LDOS) on the sublattice $A/B$ of the unit cell $(i,j)$.
After that, we follow the same procedure in Ref. \cite{wen} and select an area enclosed by the dashed square in Fig. \ref{position}(a). The location of the impurity is at the center of the square and is set to be the origin. Using this area (contains $257^2$ atoms in our calculation), we perform the Fourier transformation to get the FT-QPI as $\rho(\mathbf{q},\omega)=\sum_\mathbf{r}\rho(\mathbf{r},\omega)e^{i\mathbf{q}\cdot\mathbf{r}}$. The anti-symmetrized FT-QPI is calculated as $\delta\rho^-(\omega)=\sum_{\mathbf{q}\in A}\rm{Re}[\rho(\mathbf{q},\omega)-\rho(\mathbf{q},-\omega)]$,
%\begin{eqnarray}
%\delta\rho^-(\omega)=\sum_{\mathbf{q}\in A}\rm{Re}[\rho(\mathbf{q},\omega)-\rho(\mathbf{q},-\omega)],
%\end{eqnarray}
where the area $A$ is defined as $0.13\pi/a\leq |\mathbf{q}|\leq0.55\pi/a$, which is exactly the same area used in the experiment \cite{wen}. According to the HAEM theory \cite{hirschfeld1,hirschfeld2,wen}, $\delta\rho^-(\omega)$ should change sign when $\Delta_2\leq\omega\leq\Delta_1$ if the SC order parameter does not exhibit any sign reversal on the Fermi surfaces. Otherwise it will maintain the same sign when $\Delta_2\leq\omega\leq\Delta_1$ if the SC order parameter changes sign on the Fermi surfaces.

%\section{Results and discussion}
In the following, we show our calculated results and compare them with the experiment. In Fig. \ref{dos}(a), we plot the LDOS at the impurity site for the two pairing symmetries. As can be seen from the inset, in the clean system, the two pairing symmetries exhibit identical DOS close to the Fermi level, with two pairs of SC coherence peaks located at $\pm\Delta_1$ and $\pm\Delta_2$. At the impurity site, with appropriate scattering potential [$(V_1,V_2)=(6,-1)/(10,5)$ for the in$\&$out/hidden $s_{\pm}$ pairing], clear in-gap bound states show up, which are located at $\omega=\pm0.04$ and $\pm0.055$ for the in$\&$out $s_{\pm}$ and  hidden $s_{\pm}$ pairings, respectively. Furthermore, the intensity of the bound states at positive $\omega$ is much larger than that at negative $\omega$. The two-gap DOS in the clean system, as well as the location and the asymmetrical height of the impurity bound states, are all qualitatively consistent between our theoretical results and the experimental measurements (see Fig. 1 in Ref. \cite{wen}). In Figs. \ref{dos}(b) and \ref{dos}(c), we plot the difference of the FT-QPI $\delta\rho^-(\mathbf{q},\omega=0.085)=\rm{Re}[\rho(\mathbf{q},\omega=0.085)-\rho(\mathbf{q},\omega=-0.085)]$. The results of the two pairing symmetries show no qualitative difference and both agree with the experiment (see Fig. 3(a) in Ref. \cite{wen}. Here we show the results in the first BZ).

Then we come to $\delta\rho^-(\omega)$. In Fig. \ref{dos}(d) we plot the data extracted from Ref. \cite{wen}. As mentioned in Ref. \cite{wen}, the sharp peak at 4 meV is due to the impurity bound state and is unrelated to the phase-dependent analysis of QPI. Therefore they used a filtering scheme from 3 to 5.5 meV and the filtered $\delta\rho^-(\omega)$ is shown as the red curve. In Ref. \cite{wen}, they considered only the two electron bands and neglected the incipient bands. They claimed that, if the SC order parameter changes sign between the electron pockets (i.e., the $s^\pm$ pairing state defined in their paper), then $\delta\rho^-(\omega)$ will not change sign between $\Delta_1$ and $\Delta_2$, while it will change sign if the pairing state is $s^{++}$ (the SC order parameter does not change sign between the electron pockets). Since the experimental data show no sign change of $\delta\rho^-(\omega)$ between $\Delta_1$ and $\Delta_2$, therefore they concluded that there should exist a sign reversal of the SC order parameter on the electron pockets. The results of $\delta\rho^-(\omega)$ from our calculation are plotted in Figs. \ref{dos}(e) and \ref{dos}(f). The black solid curve in Fig. \ref{dos}(e) shows $\delta\rho^-(\omega)$ for the in$\&$out $s_{\pm}$ pairing and we can see that there is a sharp peak at $\omega=0.04$, which is due to the impurity bound state. To eliminate the effect of the bound state, we use a parabolic function $\delta\rho^-(\omega)=A\omega^2+B\omega$ to substitute the original one from $\omega=0.03$ to $0.06$, as has been done in the experiment, and show the filtered $\delta\rho^-(\omega)$ as the red dotted curve. Similarly, for the hidden $s_{\pm}$ pairing, since the impurity bound state is located at $|\omega|=0.055$, therefore we employ the same filtering scheme from $\omega=0.045$ to $0.075$, and the results are shown in Fig. \ref{dos}(f). We then rescale the filtered $\delta\rho^-(\omega)$ from our calculation and plot it with the experimental data in Fig. \ref{comparison}(a). Our theoretical results for the two pairings are both qualitatively consistent with the experimental data, that is, $\delta\rho^-(\omega)$ exhibits no sign change between $\Delta_1$ and $\Delta_2$. In the hidden $s_{\pm}$ pairing, the off-shell scattering process denoted by the green arrow in Fig. \ref{position}(b), which connects states with sign-reversed order parameter, contributes significantly to $\delta\rho^-(\omega)$ and makes $\delta\rho^-(\omega)$ in this case similar to that in the in$\&$out $s_{\pm}$ pairing case.
%In fact, if we look more closely, we find that the result of the hidden $s_{\pm}$ pairing (the red curve) agrees even better with the experimental data since they all show a peak structure between $\Delta_1$ and $\Delta_2$.
 Therefore, the experimental data does not exclusively imply a sign-changing order parameter on the electron Fermi surfaces. A detailed derivation of the HAEM theory in the presence of incipient bands can be found in Ref. \cite{supplementary_material}.

Refs. \cite{linscheid} and \cite{mishra} suggest that the location of the incipient bands may affect the SC pairing. In order to further elucidate the effects of the incipient bands on the QPI analysis, we then ignore them and repeat the above calculations.
%In calculating the Green's function $g(\mathbf{k},\omega)=(\omega+i0^+-A_\mathbf{k})^{-1}$, its matrix elements can be written as
%\begin{eqnarray}
%\label{gmn}
%g_{mn}(\mathbf{k},\omega)&=&\sum_l\frac{P_{\mathbf{k}ml}P_{\mathbf{k}nl}^{*}}{\omega+i0^+-\xi_{l\mathbf{k}}},
%\end{eqnarray}
%where $\xi_{l\mathbf{k}}$ is the $l$th eigenvalue of $A_\mathbf{k}$ in Eq. (\ref{h}) and $P_\mathbf{k}$ is a unitary matrix that diagonalizes $A_\mathbf{k}$.
In our band structure, the top of the incipient bands is located at about $-0.77$, then we consider only those bands satisfying $|\xi_{l\mathbf{k}}|\leq0.7$, where $\xi_{l\mathbf{k}}$ is the $l$th eigenvalue of $A_\mathbf{k}$ in Eq. (\ref{h}).
%in the summation of $l$ in Eq. (\ref{gmn}), we sum only those terms that satisfy the condition $|\xi_{l\mathbf{k}}|\leq0.7$.
In this way, the contribution from the incipient bands can be completely removed. We have verified that the DOS in the clean system calculated this way is not affected and is identical to those shown in the inset of Fig. \ref{dos}(a). Then we show $\delta\rho^-(\omega)$ in Fig. \ref{comparison}(b). Now the results are consistent with the HAEM theory. For the in$\&$out $s_{\pm}$ pairing, $\delta\rho^-(\omega)$ still shows no sign change between $\omega=\Delta_1$ and $\Delta_2$, since there still is a sign reversal of the SC order parameter on the Fermi surfaces. On the contrary, for the hidden $s_{\pm}$ pairing, since there is no longer the scattering process that can connect the sign-reversed order parameter as denoted by the green arrow in Fig. \ref{position}(b), $\delta\rho^-(\omega)$ exhibits a sign change between $\omega=\Delta_1$ and $\Delta_2$. A detailed evolution of $\delta\rho^-(\omega)$ with respect to the location of the incipient bands can be found in Ref. \cite{supplementary_material}.

In summary, we have investigated the momentum-integrated QPI in the FeSe-based superconductors, by taking the incipient bands into consideration. We found that, if there is SC pairing on the incipient bands, then special caution has to be taken when interpreting the pairing symmetry from the experimental data. For example, naively people may expect that the in-gap bound states induced by nonmagnetic impurities should suggest a sign-reversing order parameter on the Fermi surfaces, while our theoretical calculation indicates that this is not the case \cite{gaoy_fese,gaoy3}. In addition, the HAEM theory proposed in Refs. \cite{hirschfeld1} and \cite{hirschfeld2}, which has been used to process the experimental data in Ref. \cite{wen}, may be unable to determine the pairing symmetry in these materials. In this work, the strength of the scattering potential ($V_1,V_2$) is chosen so as to fit the location and asymmetrical height of the bound states observed in experiment. We have also verified that, for other scattering potentials, for example, if we set the interorbital scattering potential $V_2$ to be zero, the main conclusions still hold \cite{supplementary_material}. The comparison between our results and the experimental data implies that, the QPI measurement cannot distinguish the hybridization-induced in$\&$out $s_{\pm}$ pairing from the strong-coupling-spin-fluctuation-induced hidden $s_{\pm}$ pairing. Finally we would like to comment on the spin resonance observed in INS. For example, in Refs. \cite{Boothroyd} and \cite{zhaoj1}, the energy of the spin resonance is at 21 meV. However in Li$_{1-x}$Fe$_x$OHFe$_{1-y}$Se, $2\Delta_2\approx17$ meV, as determined by the STM data in Ref. \cite{wen}. Therefore the spin resonance energy is actually above $2\Delta_2$ and this can happen even if the SC order parameter preserves its sign on the Fermi surfaces \cite{kontani2,kontani3,kontani4}. Therefore, the sign of the SC order parameter in the FeSe-based superconductors is far from settled.

This work is supported by the Natural Science Foundation from Jiangsu Province of China (Grant No. BK20160094, Y.G.), the Start-up Foundation from South China Normal University (T.Z.) and NSFC (Grant No. 11574134, Q.H.W.).

\end{document}